\documentclass[a4,12pt,epsf]{article}
 
\begin{document}
\begin{center}{\Large \bf
Lepton Flavor Violating tau decay by R-Parity Violation\\ in a Family Symmetry
\footnote{Talk given at Summer Institute 2007, Aug. 3-10, 2007, Fuji-Yoshida, Japan.} 
\vspace{6pt}\\
}
\end{center}
 
\begin{center}
Yuji Kajiyama
\footnote{yuji.kajiyama@kbfi.ee}
\vspace{6pt}\\
 
{\it
National Institute of Chemical Physics and Biophysics,
Ravala 10, Tallinn 10143, Estonia
}
\end{center}
\begin{abstract}
In this talk, we investigate Lepton Flavor Violating (LFV) tau decay 
by R-Parity Violating (RPV) operators controlled by 
a non-Abelian discrete $Q_6$ family symmetry. We assume 
that only a family symmetry determines the whole flavor structure of a model, and 
the model indicates specific predictions of LFV tau decay processes by RPV operators.
We predict $BR(\tau \to 3e)/BR(\tau \to 3\mu)
\sim 4 m_{\mu}^2/m_{\tau}^2$ in a $Q_6$ family symmetric model.  
\end{abstract}
 
\section{Introduction}
Mass of matter particles and mixing angles between generations have been precisely measured. 
Mass in the quark and charged lepton sector has hierarchical structure, but only mass-squared  differences are known in the neutrino sector. As for the mixing angles, quark sector has 
small, and lepton sector has large (nearly tribimaximal) mixing angles. The Standard Model (SM) or 
the Minimal Supersymmetric Standard Model (MSSM) explain these structure of 
masses and mixings because these models contain more parameters than observables in the 
Yukawa sector. 
Therefore they can not predict the structure. Family symmetry (flavor symmetry) is the 
most promising approach to determine the Yukawa structure and predict masses and mixings. 
In general, family symmetry determines the form of the Yukawa couplings and other 
interactions such as R-Parity Violating (RPV) operators, and it reduces the 
number of parameters of the model. 
Then such models can predict masses and mixings of particles, and other processes by the symmetry. 
In this talk we consider a supersymmetric standard model with non-Abelian discrete $Q_6$ 
family symmetry, and 
obtain some predictions of Lepton Flavor Violating (LFV) tau decay processes caused by RPV operators 
which are controlled by the family symmetry \cite{kaji}. 
  
\section{The Model}
$Q_6$ group is a non-Abelian discrete subgroup of $SU(2)$ \cite{babukubo}, and it has two two-dimensional and 
four one-dimensional irreducible representations (irreps.).
In our model, three generations of matter fields are embedded into  ${\bf 2}+{\bf 1}$ dimensional 
irreps. of $Q_6$ group in a definite way. 
Moreover, both up and down type Higgs fields also 
have three generations which are embedded into irreps. in similar way, and they construct 
$Q_6$ invariant Yukawa 
interactions which give fermion mass matrices. 
After the electroweak symmetry breaking, mass matrix of the charged lepton sector is
\begin{eqnarray}
M_e &=&m_{\tau} \left(\begin{array}{ccc}
- A& 
A
&  B \\
A & 
  A &
B  \\
C  & 
C& 
0 \\
\end{array}\right)
\label{me}
\end{eqnarray}
with $A=0.04206,B=0.7057,C=0.0002893$, and $m_{\tau}$ is tau lepton mass.
One finds that diagonalization matrices $U_{eL}$ and $ U_{eR}$ are approximately 
written
as
\begin{eqnarray}
U_{eL}&\simeq&\left( \begin{array}{ccc} 
\epsilon_e & 1/\sqrt{2}
&1/\sqrt{2} \\
-\epsilon_e & -1/\sqrt{2}
&1/\sqrt{2} \\
1 & -\sqrt{2} \epsilon_e &0\\
\end{array} \right),~ 
U_{eR}\simeq \left( \begin{array}{ccc} 
0&-1&0 \\
1 &0&\epsilon_{\mu} \\
-\epsilon_{\mu} &0 & 1 \\
\end{array} \right)
, \label{UeLR}
\end{eqnarray}
where terms of ${\cal O}(\epsilon^2)$ are neglected, 
and small parameters $\epsilon_e,\epsilon_{\mu}$ are defined as
\begin{eqnarray}
\epsilon_e=\frac{m_e}{\sqrt{2}m_{\mu}}=3.42 \times 10^{-3},~
\epsilon_{\mu}=\frac{m_{\mu}}{m_{\tau}}=5.94 \times 10^{-2}.
\label{epsilon}
\end{eqnarray}
The upper-right $2 \times 2$ block of $U_{eL}$ is the origin of 
maximal mixing of the atmospheric neutrino oscillation.  

As for the neutrino sector, we assume that
a see-saw mechanism takes place.
However, we do not present the details of the neutrino sector here because 
there is no need to know it in the following analysis. We obtain some specific predictions of our model:
(i) inverted mass hierarchy $m_{\nu_3}<m_{\nu_1},m_{\nu_2}$, (ii) $|U_{e,3}|\simeq \epsilon_e$. 
See Refs. \cite{kajiyama,kubo} for details.

\section{R-Parity Violation and LFV tau decay}
In the MSSM case, RPV trilinear interactions contain 45 couplings. However
in the $Q_6$ model, the form of RPV interactions are fixed 
and the number of the couplings are reduced 
by the symmetry. The RPV trilinear operators are    
\begin{eqnarray}
W_{\not R}=\lambda L_3 L_I E_I^c+\lambda_1'  L_I (i \sigma^2)_{IJ}Q_3 D_J^c
+\lambda_2'  L_I (\sigma^1)_{IJ}Q_J D_3^c,
\label{rparity}
\end{eqnarray}
which contain only three couplings. The Baryon number violating interactions $U^c D^c D^c$ are 
forbidden by the family symmetry. Since these three couplings $\lambda,\lambda'_{1,2}$ can 
generate FCNC processes, they are constrained by experiments: 
\begin{eqnarray}
\lambda &< &1.4 \times 10^{-2}~~({\mbox{from}}~\mu \to eee)
\label{lam}\\
\lambda'_{1,2}&<&3.1 \times 10^{-3}~~({\mbox{from}}~K^0-\bar K^0 {\mbox{mixing}}).
\label{lam'}
\end{eqnarray}
These constraints are somewhat relaxed than those of the MSSM case. 
This comes from the family symmetry. Since eq.(\ref{rparity}) is 
written in flavor basis, mixing matrices eq.(\ref{UeLR}) of lepton sector and those of 
quark sector appear when the fields are transformed into mass eigenstates, and these 
mixing matrices effectively work as additional suppression factor. For instance, 
vertex of $\tau_L-\tilde \nu_{3L}-e_R$ in mass eigenstate from the $\lambda$-term 
in eq.(\ref{rparity}) is
\begin{eqnarray}
-i\lambda \sum_{J=1,2}(U_{eR}^{\dag})_{1J}(U_{eL})_{J3}=i\lambda \epsilon_e,
\end{eqnarray}  
where we have assumed that scalar mass matrix is diagonal and degenerated to forbid 
SUSY FCNC from soft SUSY breaking terms \cite{kajiyama}. \\

Now we will concentrate on the LFV decay processes $\ell^-_m \to \ell^-_i \ell^-_j \ell^+_k$. 
The operator $\lambda L_3 L_I E_I^c$ 
generates the decays $\ell_m^- \to \ell_i^- \ell_j^- \ell_k^+$ at tree level when $\lambda \neq 0$. 
The other two operators in Eq.(\ref{rparity}), $\lambda'_{1,2}LQD^c$, also generate the similar decay processes at one loop level through photon penguin diagrams, but we found that the bounds on 
$\lambda'_{1,2}$ are stronger than that on $\lambda$ in eqs.(\ref{lam}) and (\ref{lam'}). 
So we neglect contributions from $\lambda'_{1,2}$. Moreover, 
since there are three generations of both up and down type Higgs doublet in this model, 
LFV processes mediated by the neutral Higgs bosons are generated at tree level.
However, the branching ratio of the $\mu \to eee$ from these effects is $BR \sim 10^{-16}$ 
when the neutral Higgs boson mass is $100$ GeV 
because of the smallness of the Yukawa couplings and mixing matrices as extra suppression factor. 
So, these contributions are also negligible 
compared to those from $\lambda LLE^c$ couplings unless $\lambda<10^{-3}$. 
Therefore, we 
approximate that $\ell_m^- \to \ell_i^- \ell_j^- \ell_k^+$ processes are induced at tree level 
only by 
$\lambda$. In this approximation, the ratios of these processes are independent of $\lambda$, 
but depend on the mixing matrices $U_{eL(R)}$ which reflect the flavor structure of the 
model. Therefore we find some predictions of LFV decays $\ell_m^- \to \ell_i^- \ell_j^- \ell_k^+$ in our model:

 \begin{eqnarray}
\frac{BR(\tau \to eee)}{BR(\tau \to \mu \mu \mu)}
&\simeq& \frac{4 \epsilon_{\mu}^2}{1 +\epsilon_{\mu}^2}=0.014, \\
\frac{BR(\tau \to \mu\mu e)}{BR(\tau \to \mu \mu \mu)}
&\simeq& \frac{1-\epsilon_{\mu}^2+2 \epsilon_e^2}{1+\epsilon_{\mu}^2-2 \epsilon_e^2}
=0.99,~~\mbox{etc.}
\end{eqnarray}
Also, when $\lambda \sim10^{-2}$, the $\tau \to 3 \mu$ process has the largest Branching Ratio 
$BR(\tau \to 3 \mu)\sim 8 \times 10^{-9}$, and  
this can be testable at future experiments such as super B factory and LHC \footnote{
For recent results of LFV tau decay into three leptons at Belle, see ref.\cite{tau}. }.  
Therefore if $\tau \to 3 \mu$ is observed with branching ratio $\sim 10^{-9}$, $\tau \to \mu \mu e$ will also be observed almost at the same time. 
\section{Conclusions} 
We have considered phenomenology of R-parity violating interactions in non-Abelian discrete 
$Q_6$ family symmetric model. The family symmetry fixes the form of RPV operators as well as 
Yukawa interactions and mixing matrices. Since whole flavor structure is determined only 
by the family symmetry, lepton flavor violating processes induced by RPV operators are 
predicted as  $BR(\tau \to 3e)/BR(\tau \to 3 \mu)\simeq 0.014$, which reflects the 
family symmetry. Future experiments such as super B factory and LHC can test our predictions.

\vspace*{12pt}
\noindent
{\bf Acknowledgement}
\vspace*{6pt}
 
\noindent
This work is supported by the ESF grant No. 6190 and postdoc
contract 01-JD/06.              
 


\begin{thebibliography}{99}
\bibitem{kaji} 
Y. Kajiyama, JHEP, {\bf 0704}, 007 (2007).

\bibitem{babukubo}
K. S. Babu and J. Kubo, Phys. Rev. {\bf D71}, 
056006 (2005).

\bibitem{kajiyama}
E. Itou, Y. Kajiyama and J. Kubo, Nucl. Phys. {\bf B743}, 74 (2006). 

\bibitem{kubo}
J.~Kubo, A.~Mondrag\'on, M.~Mondrag\'on and
E.~Rodr\' iguez-J\' auregui,  Prog. Theor. Phys.
{\bf 109}, 795 (2003); Erratum-ibid.{\bf 114}, 287 (2005);
J.~Kubo, Phys. Lett. {\bf B578}, 156 (2004);
Erratum-ibid. {\bf B619}, 387 (2005); 
Y. Kajiyama, J. Kubo and H. Okada, Phys. Rev. {\bf D75}, 033001 (2007).

\bibitem{tau}
K. Abe {\em et al.} [Belle Collaboration], 
arXiv:0708.3272 [hep-ex].
\end{thebibliography}
\end{document}